\documentclass{article}
\usepackage[margin=1in]{geometry}
\usepackage{stackengine}

\usepackage{algorithmicx}
\usepackage[section]{algorithm}
\usepackage{algpseudocode}
\usepackage{footnote}
\usepackage{authblk}
\usepackage{color}
\usepackage{enumitem}
\usepackage{hyperref}
\usepackage{dblfloatfix}
\usepackage{url,amsmath,setspace,amssymb,tikz,graphicx}

\newcommand{\hash}{\textsf{hash}}
\newcommand{\MerkleRoot}{\textsf{MerkleRoot}}
\newcommand{\Sign}{\textsf{Sign}}
\newcommand{\Key}{\textrm{k}}
\newcommand{\Enc}{\textsf{Enc}}
\newcommand{\Dec}{\textsf{Dec}}
\newcommand{\Data}{\textsf{Data}}
\newcommand{\offchain}{\textsf{offchain}}
\newcommand{\myname}{BlockMarkchain}
\newcommand{\PartialData}{\textsf{Proof}}
\newcommand{\C}{C}
\usepackage{subcaption}

\begin{document}
	\onecolumn
	
	\title{\bf \myname: A Secure Decentralized Data Market\\ with a Constant Load on the Blockchain}
	\date{}
	\author[1]{\sl Hamidreza Ehteram\thanks{These two authors contributed equally.}\thanks{\href{mailto:ehteram.hamidreza@ee.sharif.edu}{ehteram.hamidreza@ee.sharif.edu}}}
	\author[1,2]{\sl Mohammad Taha Toghani$^*$\thanks{\href{mailto:mttoghani@rice.edu}{mttoghani@rice.edu}}}
	\author[3]{\sl Mohammad Ali Maddah-Ali\thanks{\href{mailto:mohammad.maddahali@nokia-bell-labs.com}{mohammad.maddahali@nokia-bell-labs.com}}}
	\affil[1]{Department of Electrical Engineering, Sharif University of Technology, Tehran, Iran}
	\affil[2]{Department of Electrical and Computer Engineering, Rice University, Houston, TX, USA}
	\affil[3]{Nokia Bell Labs, Holmdel, NJ, USA}
	\maketitle
	\vskip 0.3in
	\thispagestyle{empty}
	
	\begin{abstract} \noindent
				In this paper, we develop \myname, as a secure data market place, where individual data sellers can exchange certified data with buyers, in a secure environment, without any mutual trust among the parties, and without trusting on a third party, as a mediator. To develop this platform, we rely on a smart contract, deployed on a secure public blockchain. The main challenges here are to verify the validity of data and to prevent malicious behavior of the parties, while preserving the privacy of the data and taking into account the limited computing and storage resources available on the blockchain. In \myname, the buyer has the option to dispute the honesty of the seller and prove the invalidity of the data to the smart contract. The smart contract evaluates the buyer's claim and punishes the dishonest party by forfeiting his/her deposit in favor of the honest party. \myname \space enjoys several salient features including (i) the certified data has never been revealed on the public blockchain, (ii) the size of data posted on the blockchain, the load of computation on the blockchain, and the cost of communication with the blockchain is constant and negligible, and (iii) the computation cost of verifications on the parties is not expensive. 
	\end{abstract}

	\section{Introduction}\label{sec:introduction}
	These days, having access to the massive datasets on a subject becomes one of the key elements for a research or business initiative to be successful. Large companies are willing to spend a considerable amount of money to collect and process data. This motivates some organizations to develop their business as mediators for collecting and selling data to others. Still, a major amount of valuable and expensive data is lost or remained unused. For example, every day thousands of medical records, in the form of reports of medical diagnosis, treatments, test results, MRI, and X-Ray images are generated, with considerable cost,  and then lost or forgotten after treatment. Those records, if collected, can significantly facilitate and accelerate medical and pharmaceutical research and treatment.
	
	One major reason that those data remained unused is that there is no easy-to-use popular and secure platform, as a data market place, that allows individuals to present and sell their data directly, and assures them that they will benefit from that.
	
	In literature, developing such a platform is known as \emph{fair exchange problem}~\cite{pagnia2003fair}. It has been shown that there is no solution for the fair exchange problem without any trusted third party~\cite{pagnia1999impossibility}. We need a third party as a mediator between the data sellers and the buyers, to enforce the parties to fulfill their commitments, prevent malicious behavior, evaluate the validity of data, and manage disputes. The challenge is that the mediator can exploit the situation, and ask for unreasonable commission fees. He/She may lie about the real values of data, sell the data without the owner’s awareness and permission, or abuse the data for some unintended purposes. Facebook–Cambridge Analytica data scandal is only one example of those misbehaviors~\cite{cadwalladr2018facebook}.
	
	Recently, it has been shown that a smart contract, deployed on a public blockchain, can be used as the third party. Since smart contract on a blockchain is transparent, immutable, and verifiable, it does not have many disadvantages of regular mediators, such as deviating from the protocol and dishonesty. Of course, transparency can be a disadvantage too, because it may violate data privacy. This motivates~\cite{wiki2018zkcp} using zero-knowledge proofs~\cite{rosen2006zero}, in order to be able to publicly verify the honesty of the parties without revealing the data itself. However, using zero-knowledge proofs cause a huge computation burden on the parties. In this paper, our objective is to develop a blockchain-oriented solution for a data market place, with minimum communication, computational, and storage overhead on the blockchain and the parties.
	
	\begin{figure*}[btp]
		\centering
		\begin{subfigure}{0.45\textwidth}
			\centering
			\includegraphics[width=\linewidth,height=7.5cm,keepaspectratio]{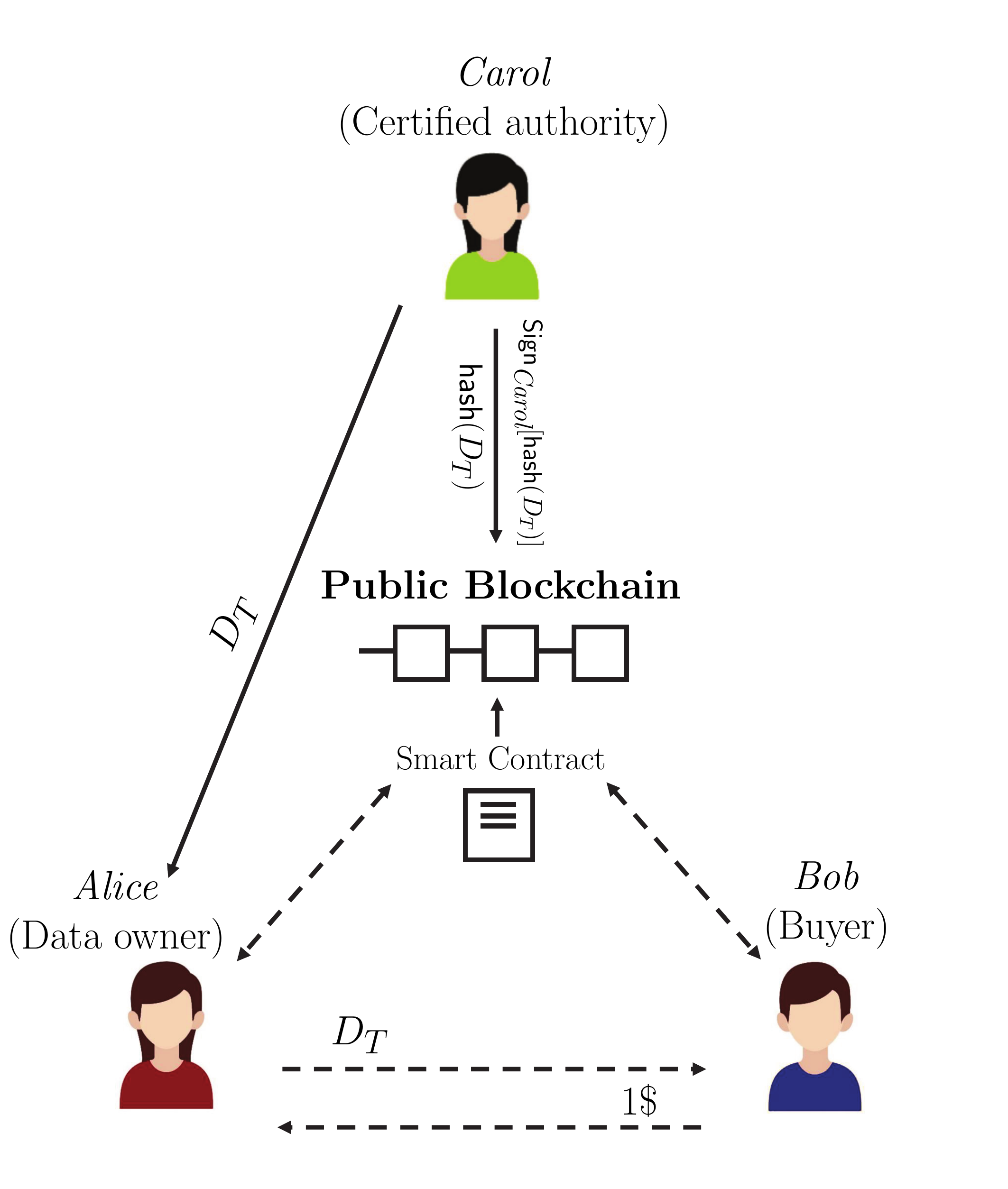}
			\caption{\emph{Alice} owns a data, denoted by $D_T$. \emph{Bob} has access to $\hash(D_T)$ which is signed by \emph{Carol} and he desires to pay say $1\$$, for $D_T$ to purchase it.}
			\label{fig_Problem}
		\end{subfigure}
		\hfill
		\begin{subfigure}{0.5\textwidth}
			\centering
			\vskip 0.5in
			\includegraphics[width=\linewidth]{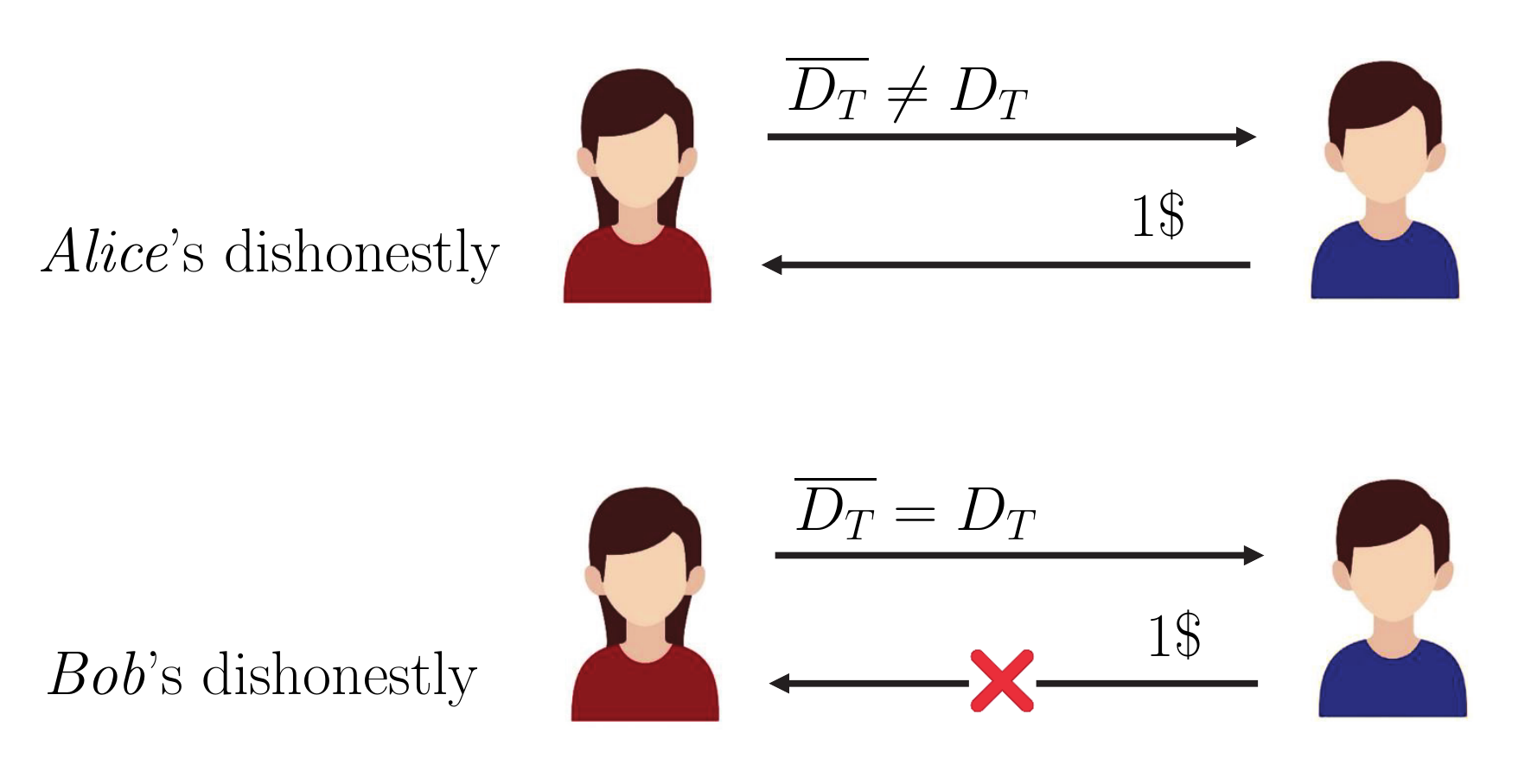}
			\vskip 0.7in
			\caption{\emph{Alice}'s dishonestly: Sending a fake data and taking money from \emph{Bob}. \space \emph{Bob}'s dishonestly: Access to the real data and refusal to pay for it.}
			\label{fig_Challenges}
		\end{subfigure}%
		\caption{Problem Statement}
	\end{figure*}
	
	We focus on a scenario, where a data seller, named \emph{Alice}, owns some data, denoted by $D_T$, which is stored on a data storage (see Figure~\ref{fig_Problem}). The data, when generated, is certified by \emph{Carol}, as a trusted individual (e.g., a medical doctor). In particular, $\hash(D_T)$, for some cryptographic hash function $\hash$, has been signed by \emph{Carol} and posted on the blockchain (we will explain these steps in details in Section~\ref{data_market}).  Signing the data by an authenticated party, \emph{Carol}, prevents generating fake data. It also specifies the possession of data by its true owner, \emph{Alice}. The data buyer, \emph{Bob}, wants to buy this data, through a smart contract, deployed on the blockchain.
	
	The proposed platform must resolve the following challenges (see  Figure~\ref{fig_Challenges})
	
	\begin{enumerate}
		
		\item \textbf{No Trusted Environment:} In the trading process, we cannot presume the seller \emph{Alice} or the buyer \emph{Bob} is trusted. We only assume the data authenticator \emph{Carol} is trusted, and she verifies validity of data $D_T$ before signing $\hash( D_T)$ and posting it on the blockchain. Let's explicate \emph{Alice} and \emph{Bob}'s dishonesty:
		\begin{enumerate}
			\item \textbf{Resistance to \emph{Alice}'s dishonesty:} If \emph{Alice} does not send valid data $D_T$ to \emph{Bob}, no money should be transfered from \emph{Bob} to \emph{Alice} and \emph{Alice} must be punished for her dishonesty.
			
			\item \textbf{Resistance to \emph{Bob}'s dishonesty:} If \emph{Bob} receives the valid data, he must not be able to deny its veracity and the agreed price must be paid to \emph{Alice}. In other words, \emph{Bob} must not be able to refrain the payment after receiving the valid data.
		\end{enumerate}
		
		\item \textbf{Privacy:} No part of the data should be revealed to anyone other than \emph{Bob}. In other words, in the process of trading data, no part of the data should be uploaded to the blockchain. Even if \emph{Bob} has a valid dispute, no part of valid data is revealed. Of course, this does not include the case where \emph{Alice} reveals the data to the public, or the case where \emph{Bob} does so after paying for it.
		
		\item \textbf{Low resource over-head on Blockchain:} With current technologies, storage and computation on the blockchain is very expensive. For a blockchain-oriented platform to work in the real world, we need to be very cautious about the computational, storage, and communication overhead that the platform imposes on the blockchain.
		
	\end{enumerate}
	
	In~\cite{ehteram2018datamarket}, posted on Github on June 20, 2018, we propose a first version of our solution, named \emph{Blockchain-based Data Market}, where we avoid any computationally-heavy cryptographic solutions such as zero-knowledge proofs. This platform is such that not only computation cost on parties is not expensive, but also computation and storage cost on the smart contract is small. In more detail, in the primary version when parties behave according to the protocol, the needed computation and storage cost to run the smart contract is constant and negligible. On the other hand, when a party deviates from the protocol and behaves dishonestly, the cost of disputation proof on the smart contract is in the order of $O(\log(N))$ where $N$ denotes the size of the primary data. More recently FairSwap~\cite{dziembowski2018fairswap} (improved in \cite{eckey2019optiSwap}) also proposes a scheme to solve the fair exchange problem. In their platform also the data size in a disputation on the smart contract is in order of $O(\log(N))$. However, the proposed scheme \cite{ehteram2018datamarket} the data transferred off-chain has a smaller size than FairSwap. In this paper, we present the modified version of Blockchain-based Data Market,\space \emph{\myname}, where we further reduce the computation and storage cost of disputation on the smart contract from $O(\log(N))$ to a constant $O(1)$.
	
	Alternative approaches, base on game theory, have been proposed in~\cite{asgaonkar2018solving,bitbay2018double}. In those solutions, each party at first commits a deposit on the smart contract. If one of the parties behave maliciously, \emph{both} parties will be punished and lose their deposit. This motivates the parties to behave honestly in trade. In the schemes of~\cite{asgaonkar2018solving,bitbay2018double}, the malicious party is not detected, and thus those schemes do not work if one party is willing to harm the other, at the cost of damaging itself. On the contrary, in \myname, the platform detects the wrongdoer and only punishes him/her.

	The rest of the paper is organized as follows. In Section~\ref{Background}, we will review the concepts of blockchain and smart contracts.  In Section~\ref{data_market}, we describe problem setting. In Section~\ref{data_market_platform}, we introduce \myname \space platform. In Sections~\ref{optimization2} and \ref{optimization3}, we further improve the proposed solution in terms of the size of uploaded data and computation on the blockchain. In Section~\ref{sec_Conclusion}, we conclude.

	\textbf{Notations:} $\hash$ denotes a cryptographic hash function, which is collision, preimage, and second-preimage resistant (see \cite{rogaway2004cryptographic} for the definitions). $\Sign_{\emph{A}}[X]$ is equal to the sign of $X$ by $\emph{A}$ signature. $\Enc_\Key(X)$ and $\Dec_\Key(X)$ denote the encrypt and decrypt version of $X$ by key $\Key$ using a secure cryptography function so $\Dec_\Key(\Enc_\Key(X)) = X$. For a function $\textsf{g}$ and a number $X$, $\C(\textsf{g}(X)) \in \mathbb{R}_+$ denotes computational complexity of calculating $\textsf{g}(X)$. $\overline{X}$ denotes a version of $X$ that one claims that it is equal to $X$. $X||Y$ denotes the concatenation of $X$ and $Y$. For $M\in \mathbb{N}$, $[M]=\{1,\ldots,M \}$. 
	
	\section{Background}
	In this section, we review blockchains and smart contracts, as the fundamental decentralized tools we use to develop our data market.
	\label{Background}
	\subsection{Blockchain}	
	In 2008, \emph{Bitcoin} was presented as a peer-to-peer decentralized cash network~\cite{nakamoto2008bitcoin}. Unlike conventional banking networks, which is based on a trusted entity (e.g., a bank) to maintain the \emph{ledger},  Bitcoin network relies on some volunteers, called \emph{miners}, to develop a public ledger of validated transactions with authenticated signatures, and to prevent fraud and double-spending. The public ledger is formed as an ordered sequence of blocks, named as \emph{blockchain}, where each block contains some transactions. Every miner has a copy of the blockchain. Miners compete to generate a new block to be added to the blockchain. Every 10 minutes on average, a new block, generated by one of the miners, will be the winner, and is broadcasted to the network. Every other miner receives this block and inspects it. If it is valid, the miner will add it to the current blockchain; otherwise, it will be discarded. The competition is based on \emph{proof of work}. In this strategy, each miner needs to solve a hash-based puzzle. In this puzzle, each miner needs to change a \emph{nonce} field in the header of the block such that the hash of the header has a specific property. The miner who generates a block with a list of valid transactions, and finds the nonce faster than the others is the winner.  
    A block reward and some transaction fees, in a cryptocurrency called Bitcoin,  have been allotted to the winner. This reward can be spent in the subsequent blocks. 

	In this process, all miners have a copy of the blockchain, where blocks in those copies become eventually consistent. It is shown that unless a major fraction of the processing power in the network is controlled by an adversary, the network is secure. 

	The fact that we can have a decentralized trusted network, without a central management,  that can maintain a database is considered as a revolutionary achievement. Central management is prone to corruption, abuse of information, intimidation,  etc.  Blockchain technology leads to new platforms for different applications where the role of central management is replaced by an immutable and transparent blockchain.

	\subsection{Smart Contract}
	Blockchain technology enables another important capability, called \emph{smart contract}. A smart contract is a computer program that is deployed on the blockchain. A user can interact with this program by issuing a transaction to the address of the smart contract. When such a transaction is received by a miner, it will run the smart contract with the transaction as the input and updates the account of that smart contract as the output. 

	Since smart contracts are deployed on the blockchain, it is transparent and immutable, anyone can read every single line of the code, observe and verify the inputs and outputs. This will expand the application of blockchain to a wide variety the cases, and allows us to develop alternative solutions for the scenarios which have been designed and managed in a central manner. Bitcoin, in the form of the locking script in the transactions, allows implementing smart contracts. However, it scripting language is not Turing complete, and thus its scope of applications is limited \cite{wiki2019script}.  The constraints of writing the code in Bitcoin protocol motivate developers to build protocols in which more sophisticated smart contracts can be implemented. In 2013, Ethereum protocol~\cite{buterin2013next} was created as a convenient platform to compose smart contracts. For example, Stroj smart contract~\cite{storj2018storj} is implemented to develop a decentralized storage platform on Ethereum. 

	Recall that the result of running a smart contract for input is verified by all miners. To do so, all miners will run the smart contract by themselves. As a result, we have to keep the computation complexity of the smart contract to be very limited. This will be controlled by the cost that miners will ask to run the smart contract.

	\section{Data Market: Problem Description}\label{data_market}
	In this section, we describe the problem statement, the requirements, and also the trust model for the data market.

	\subsection{Problem Statement}\label{problem_setting}

	We consider a scenario, where \emph{Alice} owns some data, denoted by $D_T$, and intents to transfer it to \emph{Bob} as a person who wants to buy the data for a price, which has been agreed upon, denoted by $C_{\textrm{Target}}$.
	All parties have access to a public and transparent blockchain, and a smart contract deployed on it to facilitate the exchange. Data $D_T$ is certified by \emph{Carol}, who is a trusted person. For example, \emph{Carol} can be a doctor, who witnesses the generation of the data. To certify the data, \emph{Carol} signs $\hash(D_T)$, and posts it and $\hash(D_T)$ on the blockchain, in an interaction with the smart contract. Beyond that, \emph{Carol} will not keep any record of the data and will not intervene in the process of trading the data.

	We also assume \emph{Alice} and \emph{Bob} deposit some values denoted by $C_{\textrm{deposit}A}$ and $C_{\textrm{deposit}B}$ respectively, on the smart contract. If \emph{Alice} is dishonest in the process of exchange, and sends incorrect data to \emph{Bob}, the smart contract should transfer $C_{\textrm{deposit}A}$  to \emph{Bob}. Similarly if \emph{Bob} is dishonest, and disputes the validity of the data, after receiving the genuine data, the smart contract should transfer $C_{\textrm{deposit}B}$ to \emph{Alice}. The problem here is how to design the steps of the trade process and the smart contract such that all of the requirements listed in the next subsection are fulfilled.

	\subsection{Requirements}\label{challenges}
	Assume in the end of the process, \emph{Bob} receives $\overline{D_T}$ as target data $D_T$ from \emph{Alice}. We need, the following conditions to be satisfied:
	
	\begin{enumerate}
		
	\item \textbf{No need to trust a third party in the trade process:}
	The platform should be such that it does not need any middle party (other than the smart contract) to moderate the trade process.

	\item \textbf{Presence of the certified person, \emph{Carol}, is not required during the trade process:}
	The platform should be such that only at the time that a record of \emph{Alice} is issued, the certified person needs to be available to sign the $\hash$ of the data and place it and $\hash(D_T)$ on the blockchain. Recall that \emph{Carol} is assumed to be honest and thus her signature on $\hash$ of the record in this platform means that the data with $\hash(D_T)$  is genuine. \emph{Carol} does not keep any record of the data, and is not available later to interact with.

	\item \textbf{\emph{Alice}'s dishonesty can be proved and punished:}
	If $\overline{D_T}$ is not equal to $D_T$,  \emph{Bob} must be able to dispute the trade and prove  \emph{Alice}'s dishonesty. In that case \emph{Alice} must not receive $C_{\textrm{Target}}$ and \emph{Bob} must also receive $C_{\textrm{deposit}A}$ from \emph{Alice}.

	\item \textbf{\emph{Bob}'s dishonesty can be proved and punished:}
	If $\overline{D_T}$ is equal to $D_T$, then \emph{Alice} must receive $C_{\textrm{Target}}$. In this case, if \emph{Bob} is  dishonest,  and disputes the validity of the data, the smart contract should be able to prove \emph{Bob}'s dishonestly and send $C_{\textrm{deposit}B}$, in addition to $C_{\textrm{Target}}$, to \emph{Alice}.

	\item \textbf{If the network is disconnected, before $\overline{D_T}$ is revealed to \emph{Bob}, none of the parties suffers any loss:}
	Let us assume that before $\overline{D_T}$ is revealed to \emph{Bob}, the network stops working. In this case, the situation should be as if the trade did not start at all. It means that \emph{Alice} does not have access to $C_{\textrm{Target}}$, \emph{Bob} does not receive data $D_T$, $C_{\textrm{deposit}A}$ is refundable to \emph{Alice}, and $C_{\textrm{deposit}B}$ and $C_{\textrm{Target}}$ to \emph{Bob}.

	\item \textbf{No need for the parties to do extensive computation:}
	The platform should be such that parties in a trade (\emph{Alice} and \emph{Bob}) don't need to execute large computations.

	\item \textbf{No need to place bulk of data on the blockchain:}
	We know that uploading data to a public blockchain costs a lot. For example, the cost of uploading $1 MB$ data in Ethereum blockchain is about $\$ 600$ based on the current price of Ether on 22 August 2019~\cite{wood2014ethereum,ethgasstation2019recommended}. 
	The platform should be designed such that it does not store bulky data on the blockchain.

	\item \textbf{No need to execute extensive computation on the blockchain:}
	As we know, all computations on a smart contract must be verified by all miners thus it costs too much if these computations are extensive. 
	Miners may refuse to mine and verify such transactions.

	\item \textbf{Privacy of the data must be preserved:}
	The platform should be such that during the trade process, the data is not revealed to the network.

	\end{enumerate}

	\subsection{The Trust Model}\label{trust_model}
	In this problem, we consider the following \emph{trust model} under which we design and improve the proposed platform:
	\begin{enumerate}
		\item We assume that \emph{Alice} and  \emph{Bob} are not trusted and may act maliciously. 
		\item We assume that \emph{Carol} is trusted. If $\hash(\overline{D_T})$ is matched with $\hash(D_T)$, which has been signed and posted by \emph{Carol} on blockchain, then $\overline{D_T}$ is genuine. 
		\item The public blockchain is secure, transparent, and immutable. The smart contract, its input, and its state (or its account), is transparent to everyone. 
	\end{enumerate}

	\section{The Data Market Platform}\label{data_market_platform}
	In previous section, we described the problem formulation and requirements. In Subsection \ref{platform_presentation}, we present \myname \space platform. In Subsection~\ref{challenges_resolving}, we prove how the proposed platform satisfies the required conditions stated in the problem. In Sections~\ref{optimization2} and~\ref{optimization3}, we further improve the proposed algorithm in terms of disclosure of the data and the cost of storage and computation on the blockchain respectively from $O(N)$ to $O(\log(N))$ and from $O(\log(N))$ to $O(1)$.
	
	\subsection{Platform Presentation}\label{platform_presentation}
\begin{figure}[btp]
		\centering
		\includegraphics[width=90mm,height=10cm,keepaspectratio]{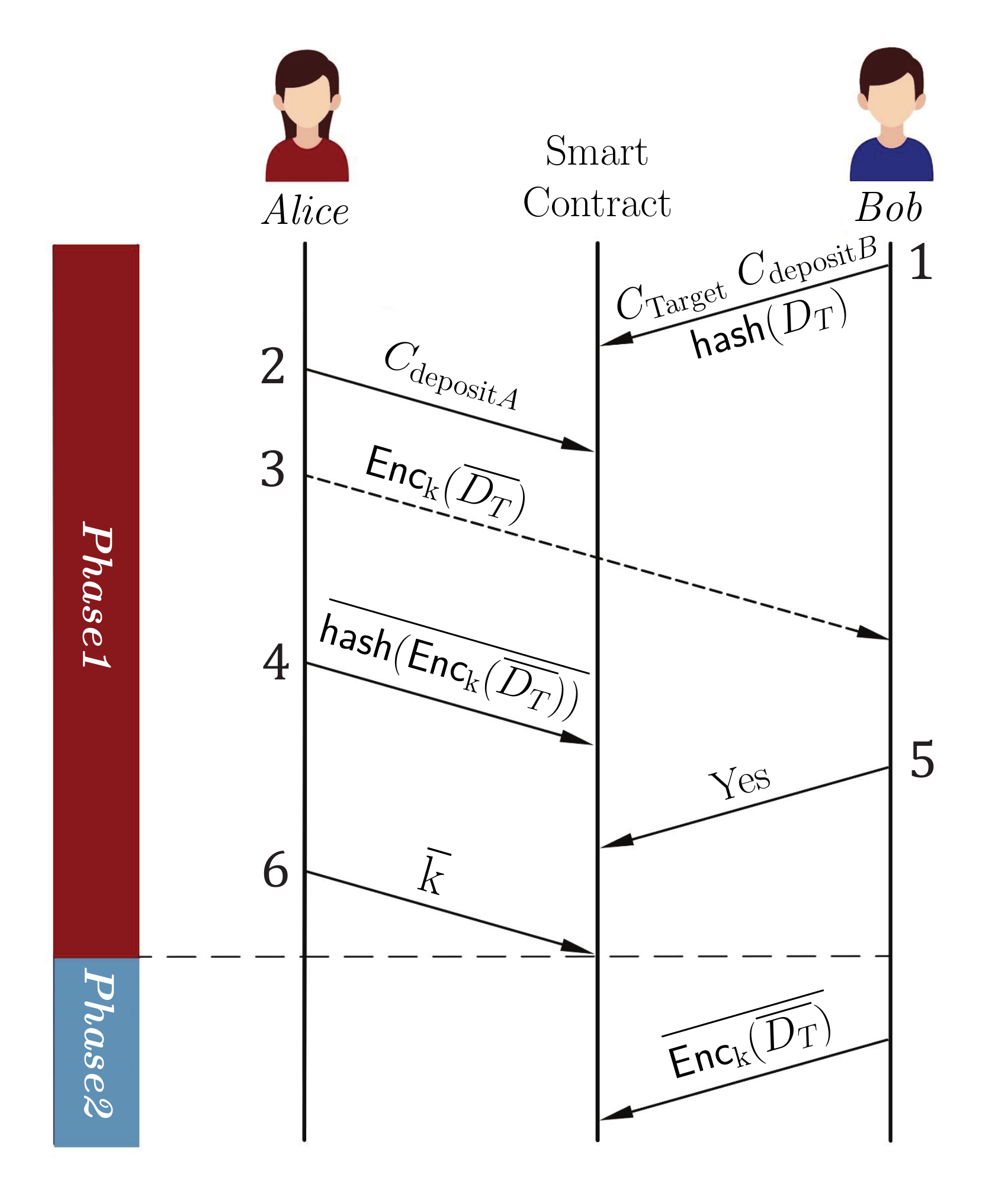}
		\caption{The messages transferred in $O(N)$-Algorithm}
		\label{primarySolution}
	\end{figure}

	In this section, we present the proposed scheme which includes two phases (Figure~\ref{primarySolution}).
	
	\subsubsection{\emph{Trading Phase} }

	In this phase, \emph{Alice} and \emph{Bob} follows Algorithm~~\ref{algorithm1} to interact with each other through a smart contract to exchange $D_T$.

	\begin{algorithm}[btp]
	\caption{ $O(N)$-Algorithm: Trading Phase}
	\label{algorithm1}
	\begin{algorithmic}[1]
\Procedure {Initialization}{}
 	\State \emph{Alice} owns data $D_T$, which is certified by  \emph{Carol}. This means \emph{Carol} has signed $\hash(D_T)$ as $\Sign_{\emph{Carol} }[\hash(D_T)]$ and posted $\hash(D_T)$ and $\Sign_{\emph{Carol} }[\hash(D_T)]$ on the smart contract. 
 	\State \emph{Alice} and \emph{Bob} agree upon price $C_{\textrm{Target}}$, and also $C_{\textrm{deposit}A}$ and $C_{\textrm{deposit}B}$. 
 	\State \emph{Alice} generates a key, denoted by \Key.	
	\EndProcedure
	\Statex
	\Procedure {Trading Phase}{}
			\State \emph{Bob} sends $\hash(D_T)$ to the smart contract, showing he has interest in buying $D_T$. 
			\State \emph{Bob} deposits $C_{\textrm{Target}}$ and $C_{\textrm{deposit}B}$ to the smart contract, nonrefundable for one day.
			\State \emph{Alice} deposits $C_{\textrm{deposit}A}$ to the smart contract, nonrefundable for one day.
			
			\Comment{If Algorithm doesn't get to Step~\ref{sendK} until the end of one day, it will go to Step~\ref{refundable}.}

			\State \emph{Alice} generates  $\Enc_\Key(\overline{D_T})$  as an  encrypted version of  $\overline{D_T}$, using key $\Key$, where $\overline{D_T}$ is the data that she claims to be equal to $D_T$. \emph{Alice} sends $\Enc_\Key(\overline{D_T})$ to \emph{Bob} using an off-chain channel (a custom P2P channel). \label{sendingencK}
	         	\State{\emph{Alice} commits $\overline{ \hash (\Enc_\Key (\overline{D_T}))  }$ to the smart contract, claiming it is indeed $\hash$ of  $\Enc_\Key(\overline{D_T})$ .} \label{commithashA}
			\State{\emph{Bob} checks if $\overline{ \hash (\Enc_\Key (\overline{D_T}))}$, posted on the smart contract is indeed equal to $\hash$ of  $\Enc_\Key(\overline{D_T})$, then sends "Yes" to the smart contract if the equality is verified. In other words, \emph{Bob} checks this equality:
			$$\hash(\Enc_\Key(\overline{D_T})) == \overline{ \hash (\Enc_\Key(\overline{D_T}))}$$
			If \emph{Bob} sends "No" to the smart contract or remains silent until the end of one day, the smart contract goes to Step~\ref{refundable}.} \label{sendYes}

			\State{After receiving "Yes" by the smart contract from \emph{Bob}, \emph{Alice} sends $\overline{\Key}$ to the smart contract, claiming it is indeed key $\Key$.}
			\label{sendK}
			\State{\emph{Bob} can check the validity of the received data in Step~\ref{sendingencK} by decrypting it using key $\overline{\Key}$ (available on the smart contract), then computing the $\hash$ of the decrypted data, and then comparing it with $\hash(D_T)$. In other words, he can check the following equality:
			$$\hash\Big(\Dec_{\overline{\Key}}\big(\Enc_\Key(\overline{D_T})\big)\Big) == \hash\big(D_T\big)$$
		If the equality is not verified by \emph{Bob}, he sends $\Enc_\Key(\overline{D_T})$ to the smart contract. Otherwise, he does not send anything to the smart contract.} \label{Bob_hash_dec}
		\If {the smart contract receives no objection from \emph{Bob} in a determined grace period (say 2 days)}
		\State {$C_{\textrm{Target}}$ is refundable to \emph{Alice}, and $C_{\textrm{deposit}A}$ and $C_{\textrm{deposit}B}$ to \emph{Alice} and \emph{Bob} respectively.}
		\State{Algorithm terminates.}
		\Else
		\State Go to Disputation Phase (Algorithm~\ref{algorithm2}).
		\EndIf
		\State{Deposits be refundable.
			
			$C_{\textrm{Target}}$ and $C_{\textrm{deposit}B}$ are refundable to \emph{Bob} and $C_{\textrm{deposit}A}$ to \emph{Alice}. Algorithm terminates.}\label{refundable}
				\EndProcedure
		\end{algorithmic}
	\end{algorithm}

			\subsubsection{\emph{Disputation Phase} }

	\begin{algorithm}[H]
		\caption{$O(N)$-Algorithm: Disputation Phase}
		\label{algorithm2}
		\begin{algorithmic}[1]
		\Procedure {Disputation Phase}{}
		\State \emph{Bob}  sends  $ \overline{\Enc_\Key(\overline{D_T})}$ to the smart contract, claiming he received it from \emph{Alice} in Step~\ref{sendingencK}  of Algorithm~\ref{algorithm1}. \label{al2s1}
		
		\Comment{ Recall that \emph{Bob} received $\Enc_\Key(\overline{D_T})$ from \emph{Alice} in Step~\ref{sendingencK} of Algorithm~\ref{algorithm1}.
}
		
\State Smart contract computes $\hash(\overline{\Enc_\Key(\overline{D_T})})$ and compares it with  $\overline{\hash(\Enc_\Key(\overline{D_T}))}$, received in  Step~\ref{commithashA} of Algorithm~\ref{algorithm1}, from \emph{Alice}. \label{Al2s2}

\If {$\hash(\overline{\Enc_\Key(\overline{D_T})}) == \overline{ \hash (\Enc_\Key(\overline{D_T}))  }$}\label{Al2s3}
\State Using key $\overline{\Key}$, the smart contract calculates $\Dec_{\overline{\Key}}(\overline{\Enc_\Key(\overline{D_T})})$  as decryption of $\overline{\Enc_\Key(\overline{D_T})}$, and then computes $\hash(\Dec_{\overline{\Key}}(\overline{\Enc_\Key(\overline{D_T})}))$.

		\Comment{ Recall that the smart contract received $\overline{ \hash (\Enc_\Key(\overline{D_T}))  }$ and $\overline{\Key}$ from \emph{Alice} in Steps ~\ref{commithashA} and~\ref{sendK}
		of Algorithm~\ref{algorithm1}.}

\If {$\hash(\Dec_{\overline{\Key}}(\overline{\Enc_\Key(\overline{D_T})}))  = = \hash(D_T)$}\label{Al2s5}
\State{go to Step~\ref{badBob}: 
 \emph{Bob} is dishonest.}
\Else
\State{ go to Step~\ref{badAlice}: 
 \emph{Alice} is dishonest.}
\EndIf\label{Al2s9}
\Else
\State{ go to Step~\ref{badBob}: 
 \emph{Bob} is dishonest.}
\EndIf

			\State{\emph{Alice} is dishonest.
				
				$C_{\textrm{Target}}$, $C_{\textrm{deposit}A}$, and $C_{\textrm{deposit}B}$ are refundable to \emph{Bob}. Algorithm terminates.}

						\label{badAlice}
						
			\State{\emph{Bob} is dishonest.
				
				$C_{\textrm{Target}}$, $C_{\textrm{deposit}A}$, and $C_{\textrm{deposit}B}$ are refundable to \emph{Alice}. Algorithm terminates. }

			\label{badBob}
\EndProcedure
		\end{algorithmic}
		
	\end{algorithm}

	\subsection{Addressing the Requirements}\label{challenges_resolving}
	Now, we will explain how the proposed algorithm addresses the challenges listed in Subsection~\ref{challenges}:
	
	\begin{enumerate}
		\item \textbf{No need to trust a third party in the trade process:}
		By carefully reviewing the proposed platform, it is obvious that does not need any middle man since the smart contract takes care of the integrity of the trade between \emph{Alice} and \emph{Bob}.

		\item \textbf{No need to presence of the certified person, \emph{Carol}, during the trade process:}
		The proposed algorithm does not need any attendance of the certified person during the trade process.

		\item \textbf{Overcoming \emph{Alice}'s dishonesty:}
		If \emph{Alice} sends another data instead of $D_T$ to \emph{Bob}, according to Step~\ref{al2s1} of Algorithm~\ref{algorithm2}, \emph{Bob} will send the encrypted version of $D_T$ ($ \overline{\Enc_\Key(\overline{D_T})}$) to the smart contract. The smart contract already has key $\overline{\Key}$ and can investigate and confirm that \emph{Alice} took a fraudulent step (see Steps \ref{Al2s5}-\ref{Al2s9} of Algorithm \ref{algorithm2}). So, in Step~\ref{badAlice} of Algorithm~\ref{algorithm2}, the smart contract gives all money ($C_{\textrm{Target}}$, $C_{\textrm{deposit}A}$, and $C_{\textrm{deposit}B}$) to \emph{Bob} as a penalty. Note that $C_{\textrm{deposit}A}$ must cover the cost of the data uploading to the blockchain by \emph{Bob} in the case that disputation happens.

		\item \textbf{Overcoming \emph{Bob}'s dishonesty:}
		\emph{Bob}'s dishonesty means that he has received $D_T$, but he disputes to take all deposits ($C_{\textrm{Target}}$, $C_{\textrm{deposit}A}$, and $C_{\textrm{deposit}B}$) by uploading another data instead of the data which has received off-chain. But, such a fraud can be detected. Recall that \emph{Bob}, in Step~\ref{sendYes} of Algorithm~\ref{algorithm1}, already has confirmed that the $\hash$ of received encrypted data is equal to the claimed $\hash$ on the smart contract in Step~\ref{commithashA} of Algorithm~\ref{algorithm1}, \Big(i.e., $ \hash(\Enc_\Key(\overline{D_T})) = \overline{ \hash (\Enc_\Key (\overline{D_T}))}$\Big). Therefore if he claims that some other data has been received through the off-chain channel, according to the second-preimage resistancy of the $\hash$ function, his data does not have the same $\hash$ value as $\overline{\hash(\Enc_\Key(\overline{D_T}))}$ to pass the condition of Step~\ref{Al2s3} of Algorithm~\ref{algorithm2}. If \emph{Bob} still claims this, then the smart contract will send his deposit ($C_{\textrm{deposit}B}$), in addition to $C_{\textrm{Target}}$, to \emph{Alice}. 
		
		\item \textbf{If the network is disconnected, before $\overline{D_T}$ is revealed to \emph{Bob}, none of the parties suffers any loss:}
		After sending key $\overline{\Key}$ to the smart contract (in Step~\ref{sendK} of Algorithm~\ref{algorithm1}), $\overline{D_T}$ is revealed to \emph{Bob}, hence if the network is disconnected before it, the smart contract sends $C_{\textrm{deposit}A}$ to \emph{Alice}, and $C_{\textrm{deposit}B}$ and $C_{\textrm{Target}}$ to \emph{Bob} (in Step~\ref{refundable} of Algorithm~\ref{algorithm1}).
		
		\item \textbf{No need for the parties to do extensive computation:}
		According to Algorithm~\ref{algorithm1}, the computational overhead on \emph{Alice} is computing the encryption of $D_T$ and the $\hash$ of  $\Enc_\Key(\overline{D_T})$ (in Steps~\ref{sendingencK} and \ref{commithashA}), and on \emph{Bob} is computing the $\hash$ of $\Enc_\Key(\overline{D_T})$, the decryption of $\Enc_\Key(\overline{D_T})$, and the $\hash$ of $\Dec_{\overline{\Key}}(\Enc_\Key(\overline{D_T}))$ (in Steps~~\ref{sendYes} and \ref{Bob_hash_dec}). Let us assume that for any data $D$, $\frac{|\Enc_\Key(D)|}{|D|}=\alpha$, $\frac{\C(\Enc_\Key(D))}{|D|}=\beta_1$, $\frac{\C(\Dec_\Key(D))}{|D|}=\beta_2$, and $\frac{\C(\hash(D))}{|D|}=\beta_3$, for some $\alpha \geq 1$ and $\beta_1, \beta_2$, and $\beta_3 > 0$. Then the computational overhead on the parties is $O(N)$, where $N=|D_T|$.

		\item \textbf{No need to place bulk of data on the blockchain:}
Reviewing~Algorithms~\ref{algorithm1}  and~\ref{algorithm2}, one can see that if the disputation phase does not happen, then the size of the data uploaded to the blockchain is constant and negligible. 
On the other hand, if disputation happens, according to Algorithm~\ref{algorithm2}, \emph{Bob} needs to upload the encrypted data that he received through the off-chain channel, to the smart contract. We know that the size of the encrypted version of the data is proportional to the size of the data, which is large, in the order of the size of the data i.e., $N$.
In the next sections, we will modify the proposed algorithm and resolve this issue. 	
	
	\item \textbf{No need to execute extensive computation on the blockchain:}
Again, one can confirm that in Algorithm~\ref{algorithm1}, the computation cost of the smart contract is very limited. However, if the disputation phase is called, then  calculating $\hash(\overline{\Enc_\Key(\overline{D_T})})$ and $\hash(\Dec_{\overline{\Key}}(\overline{\Enc_\Key(\overline{D_T})}))$ in Steps~\ref{Al2s3} and \ref{Al2s5} of Algorithm~\ref{algorithm2} require $O(N)$ computation, which is not desired. We will resolve this issue in the modified algorithm in the next sections.

		\item \textbf{Privacy of the data must be preserved:}
		We can confirm that the proposed algorithm addresses  the privacy challenges by considering two different scenarios: 
		\begin{enumerate}
			\item If the disputation does not happen,  the final data is revealed only to \emph{Bob} and never to the blockchain. Thus, it is not publicly available and the privacy of the data is preserved.
			
			\item The disputation happens because \emph{Alice} has sent a wrong data to \emph{Bob}. In that case, \emph{Bob} initiates the disputation phase and in that phase $\overline{D_T}$ will be revealed. However, $\overline{D_T}$ is not the same as $D_T$ and thus revealing it does not violate privacy. 
		\end{enumerate}

There is a concern here. In the case of disputation, if Algorithm~\ref{algorithm2} happens, even if $D_T$ and $\overline{D_T}$ are different in a bit, and then $\overline{D_T}$ will be revealed entirely.  We will resolve this issue in the modified algorithms in the next sections. 
			\end{enumerate}

\section{Platform Improvement: $O(\log(N))$-Algorithm:}\label{optimization2}

\begin{figure*}[btp]
	\centering
	\begin{subfigure}{\textwidth}
		\centering
		\includegraphics[width=0.61\linewidth]{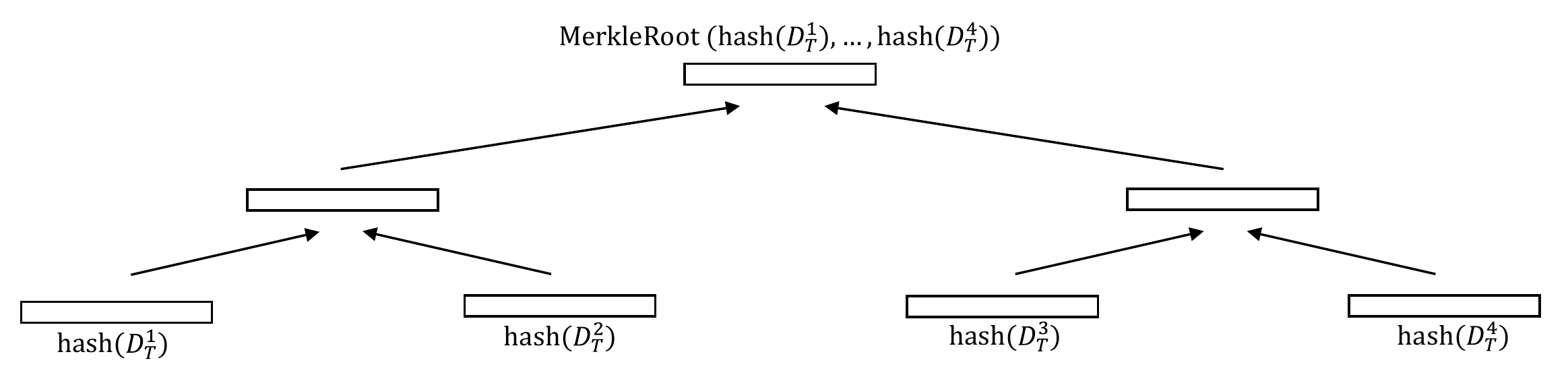}
		\caption{Data $D_T$ is split into $M\in \mathbb{N}$ chunks (say $M=4$), as $D_T = (D_T^1,\ldots,D_T^M)$.\\ \emph{Carol} uploads $\MerkleRoot(\hash(D_T^1),\ldots,\hash(D_T^M))$ and $\Sign_{\emph{Carol}}\Big[\MerkleRoot\big(\hash(D_T^1),\ldots,\hash(D_T^M)\big)\Big]$ to the smart contract.}
		\label{fp11a}
	\end{subfigure}
	\hfill
	\begin{subfigure}{\textwidth}
		\centering
		\includegraphics[width=0.7\linewidth]{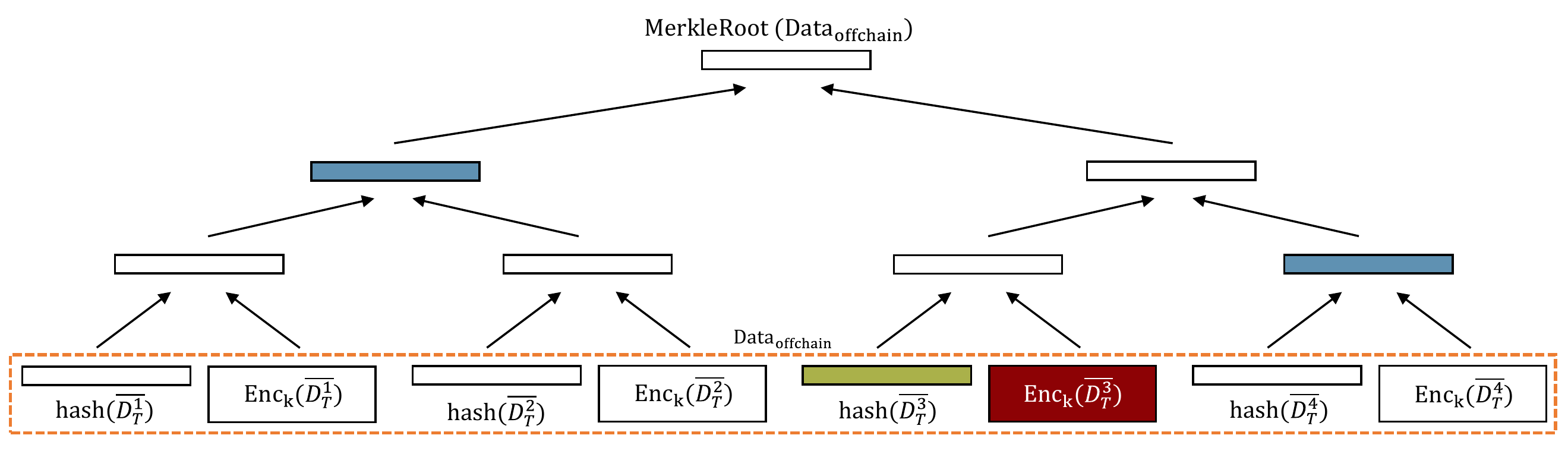}
		\caption{\emph{Alice} sends sequence $\Data_\offchain \triangleq \left(  \hash(\overline{D_T^1}), \Enc_\Key(\overline{D_T^1}),  \hash(\overline{D_T^2}), \Enc_\Key(\overline{D_T^2}), \ldots, \hash(\overline{D_T^M}), \Enc_\Key(\overline{D_T^M}) \right)$, inclosed in the orange box in the figure, to \emph{Bob} through the off-chain channel.  \emph{Alice} also  commits $\overline{\MerkleRoot(\Data_\offchain)}$ to the smart contract, claiming it is equal to $\MerkleRoot$ of  $\Data_\offchain$.\\
		If \emph{Bob} disputes the validity of a chunk of data (say $3^{rd}$ chunk), represented by red rectangle, he sends $\hash(\overline{D_T^3}) || \Enc_\Key(\overline{D_T^3})$ and its Merkle proof (shown by blue rectangles) to the smart contract.}
		\label{fp11b}
	\end{subfigure}%
	\caption{Merkle Tree for $O(\log(N))$-Algorithm}\label{fig_Optimization2}
\end{figure*}

		In this modified version of the algorithm, \emph{Carol} uses Merkle Tree~\cite{merkle1980protocols} of data $D_T$, in a certain way, described in Algorithm~\ref{algorithm1_2}, and commits the $\hash$ of the root and the signed version of it to the smart contract (see Figure~\ref{fp11a}). This approach will reduce the size of the data, uploaded to the smart contract in the disputation phase, and maximum computation load of the smart contract from $O(N)$ to $O(\log(N))$, as detailed in  Algorithm~\ref{algorithm1_2} (Figure~\ref{fig_Optimization2}). 

		\begin{algorithm}[btp]
		\caption{$O(\log(N))$-Algorithm: Trading Phase}
		\label{algorithm1_2}
		\begin{algorithmic}[1]
\Procedure {Initialization}{}
			\State \emph{Alice} owns data $D_T$, which is divided into $M \in \mathbb{N}$ chunks, each of size $L \in \mathbb{N}$ bits, as $D_T = (D_T^1,\ldots,D_T^M)$. Data $D_T$ is certified by \emph{Carol}. This means \emph{Carol} has uploaded $\MerkleRoot(\hash(D_T^1),\ldots,\hash(D_T^M))$ and $\Sign_{\emph{Carol}}\Big[\MerkleRoot\big(\hash(D_T^1),\ldots,\hash(D_T^M)\big)\Big]$ to the smart contract (see Figure~\ref{fp11a}).
			\State \emph{Alice} and \emph{Bob} agree upon price $C_{\textrm{Target}}$, and also $C_{\textrm{deposit}A}$ and $C_{\textrm{deposit}B}$.
			\State \emph{Alice} generates a key, denoted by \Key.	
\EndProcedure
	\Statex
\Procedure {Trading Phase}{} 
			\State{\emph{Bob} sends $\MerkleRoot \big(\hash(D_T^1),\ldots,\hash(D_T^M)\big)$ to the smart contract, showing his interest in buying $D_T$.}
			\State{\emph{Bob} deposits $C_{\textrm{Target}}$ and $C_{\textrm{deposit}B}$ to the smart contract, nonrefundable for one day.}
			\State{\emph{Alice} deposits $C_{\textrm{deposit}A}$ to the smart contract, nonrefundable for one day.}
			
			\Comment{If Algorithm doesn't get to Step~\ref{sendK_2} until the end of one day, it will go to Step~\ref{refundable_2}.}

						\State{\emph{Alice} generates $\Enc_\Key(\overline{D_T^m}), \forall m \in [M],$ as an  encrypted version of  $\overline{D_T^m}$, using key $\Key$, where $\overline{D_T^m}$ is the $m^{th}$ chunk of data $\overline{D_T}$ (Similar to Algorithm~\ref{algorithm1}, $\overline{D_T}$ is the data that Alice sends and claims to be $D_T$). \emph{Alice} sends $\Data_\offchain \triangleq \left(  \hash(\overline{D_T^1}), \Enc_\Key(\overline{D_T^1}),  \hash(\overline{D_T^2}), \Enc_\Key(\overline{D_T^2}), \ldots, \hash(\overline{D_T^M}), \Enc_\Key(\overline{D_T^M}) \right)$ to \emph{Bob} through the off-chain channel.}\label{sendingencK_2}

			\State{\emph{Alice} commits $\overline{\MerkleRoot(\Data_\offchain)}$ to the smart contract, claiming it is equal to $\MerkleRoot$ of  $\Data_\offchain$ (Figure~\ref{fp11b}).}\label{commithashA_2}
			\State{\emph{Bob} checks if:
				\begin{enumerate}[label=\alph*]
					\item $\MerkleRoot(\Data_\offchain) == \overline{\MerkleRoot(\Data_\offchain)}$
					\item $\MerkleRoot(\hash(\overline{D_T^1}),\ldots,\hash(\overline{D_T^M})) == \MerkleRoot(\hash(D_T^1),\ldots,\hash(D_T^M))$
				\end{enumerate}
				and sends "Yes" to the smart contract, if the above equalities hold. If \emph{Bob} sends "No" to the smart contract or remains silent until the end of one day, the smart contract goes to Step~\ref{refundable_2}}.\label{sendYes_2}
			
			\State{After receiving "Yes" by the smart contract from \emph{Bob}, \emph{Alice} sends $\overline{\Key}$ to the smart contract, claiming it is indeed key $\Key$.}\label{sendK_2}
			\State{\emph{Bob} can check validity of the received data by decrypting each encrypted chunk using key $\overline{\Key}$ (available on the smart contract), then computing the $\hash$ of each decrypted chunk, and then comparing each with the version that \emph{Alice} sent to him through the off-chain channel. In other words, he can check the following equality for all chunks:
				$$\hash\Big(\Dec_{\overline{\Key}}\big(\Enc_\Key(\overline{D_T^m})\big)\Big) == \hash\big(\overline{D_T^m}\big), m \in [M]$$
			If the above equality is not valid for at least one chunk, say chunk $w \in [M]$, \emph{Bob} sends $\hash(\overline{D_T^w}) || \Enc_\Key(\overline{D_T^w})$, and its Merkle proof for $\overline{\MerkleRoot(\Data_\offchain)}$ to the smart contract. Otherwise, he does not send anything to the smart contract.}\label{Bob_hash_Dec_2}
			\If {the smart contract receives no objection from \emph{Bob} in a determined grace period (say 2 days)}
			\State {$C_{\textrm{Target}}$ is refundable to \emph{Alice}, and $C_{\textrm{deposit}A}$ and $C_{\textrm{deposit}B}$ to \emph{Alice} and \emph{Bob} respectively.}
			\State{The algorithm terminates.}
			\Else
			\State Go to the Disputation Phase (Algorithm~\ref{algorithm2_2}).
			\EndIf
			\State{Deposits be refundable.
				
				$C_{\textrm{Target}}$ and $C_{\textrm{deposit}B}$ are refundable to \emph{Bob} and $C_{\textrm{deposit}A}$ to \emph{Alice}. Algorithm terminates.}\label{refundable_2}
				\EndProcedure
		\end{algorithmic}
	\end{algorithm}

	
	\begin{algorithm}[H]
	\caption{$O(\log(N))$-Algorithm: Disputation Phase}
	\label{algorithm2_2}
	\begin{algorithmic}[1]
	\Procedure {Disputation Phase}{}
		\State{If there is a chunk $w \in [M]$, such that $\hash\Big(\Dec_{\overline{\Key}}\big(\Enc_\Key(\overline{D_T^w})\big)\Big) \neq \hash\big(\overline{D_T^w}\big)$, \emph{Bob} sends its $\hash$ and encrypted version, i.e., $\overline{\hash(\overline{D_T^w})} || \overline{\Enc_\Key(\overline{D_T^w})}$, along with its Merkle proof for $\overline{\MerkleRoot(\Data_\offchain)}$ to the smart contract, claiming he received it from \emph{Alice} in Step~\ref{sendingencK_2} of Algorithm~\ref{algorithm1_2}. We denote these data as $\overline{\PartialData_\offchain}$.}

		\Comment{Recall that \emph{Bob} received $\Data_\offchain$ from \emph{Alice} in Step~\ref{sendingencK_2} of Algorithm~\ref{algorithm1_2}}.

		\State{Using the uploaded data by \emph{Bob}, the smart contract verifies $\overline{\PartialData_\offchain}$ for $\overline{\MerkleRoot(\Data_\offchain)}$, received in Step~\ref{commithashA_2} of Algorithm~\ref{algorithm1_2}.}
		
		\If{ $\overline{\PartialData_\offchain}$ is verified for $\overline{\MerkleRoot(\Data_\offchain)}$} \label{cost1_2}
		
		\State{Using key $\overline{\Key}$, the smart contract calculates $\Dec_{\overline{\Key}}\big(\overline{\Enc_\Key(\overline{D_T^w})}\big)$  as decryption of $\overline{\Enc_\Key(\overline{D_T^w})}$, and then computes $\hash\Big(\Dec_{\overline{\Key}}\big(\overline{\Enc_\Key(\overline{D_T^w})}\big)\Big)$.}
			
			\Comment{ Recall that the smart contract received $\overline{\MerkleRoot(\Data_\offchain)}$ and $\overline{\Key}$ from \emph{Alice} in Steps~\ref{commithashA_2} and~\ref{sendK_2}
				of Algorithm~\ref{algorithm1_2}.}
			
			\If {$\hash\Big(\Dec_{\overline{\Key}}\big(\overline{\Enc_\Key(\overline{D_T^w})}\big)\Big) == \overline{\hash(\overline{D_T^w})}$} \label{cost2_2}
			\State{go to Step~\ref{badBob_2}: 
				\emph{Bob} is dishonest.}
			\Else
			\State{ go to Step~\ref{badAlice_2}: 
				\emph{Alice} is dishonest.}
			\EndIf
			\Else
			\State{ go to Step~\ref{badBob_2}: 
				\emph{Bob} is dishonest.}
			\EndIf

		\State{\emph{Alice} is dishonest.
			
			$C_{\textrm{Target}}$, $C_{\textrm{deposit}A}$, and $C_{\textrm{deposit}B}$ are refundable to \emph{Bob}. Algorithm terminates.}

		\label{badAlice_2}
		
		\State{\emph{Bob} is dishonest.
			
			$C_{\textrm{Target}}$, $C_{\textrm{deposit}A}$, and $C_{\textrm{deposit}B}$ are refundable to \emph{Alice}. Algorithm terminates. }
		
		\label{badBob_2}
\EndProcedure
		\end{algorithmic}
	\end{algorithm}
	
	\subsection{Analysis of $O(\log(N))$-Algorithm}
	It is easy to verify that the first to fifth requirements are satisfied in this algorithm. Also, it is easy to see that the computation and storage cost of the trading phase to the smart contract is constant. Here we want to evaluate the computation load for the parties and the computation and storage cost of the disputation phase to the smart contract.
	\subsubsection{Computation Load for the Parties}
		According to Algorithm~\ref{algorithm1_2}, the computation load for \emph{Alice} is computing the encryption of each chunk of the data and $\MerkleRoot$ of $\Data_\offchain$ (Steps~\ref{sendingencK_2} and \ref{commithashA_2}), and for \emph{Bob} is computing $\MerkleRoot(\Data_\offchain)$, $\MerkleRoot(\hash(\overline{D_T^1}),\ldots,\hash(\overline{D_T^M}))$, and $\hash\Big(\Dec_{\overline{\Key}}\big(\Enc_\Key(\overline{D_T^m})\big)\Big), \forall m \in [M]$ (Steps~\ref{sendYes_2} and \ref{Bob_hash_Dec_2}). Similar to the analysis of the previous algorithm, the computation load for the parties is $O(N)$, where $|D_T|=N$.
	
	\subsubsection{Size of the Data Uploaded to the Blockchain}
As we argued, in Algorithm~\ref{algorithm1},  in the disputation phase, the size of the data uploaded to the smart contract is $O(N)$. Here we argue that in modified Algorithm~\ref{algorithm1_2} and ~\ref{algorithm2_2}, this is reduced to $O(\log(N))$. 

Let us assume $|D_T| = N$, and thus the number of chunks is equal to $M = \lceil\frac{N}{L}\rceil$. In addition, let us assume that for any data $D$, $\frac{|\Enc_\Key(D)|}{|D|}=\alpha$, for some $\alpha \geq 1$. Then the size of $\overline{\PartialData_{\offchain}}$, the uploaded data to the smart contract, in this scheme is equal to:
		$$(\log_2 M+1)h + \alpha L=(\log_2(\frac{N}{L})+1)h + \alpha L$$
		where $h$ denotes the size of the $\hash$ output. The optimum size of $L$ to minimize the uploaded cost is equal to: 
		$$L^* = \frac{h}{\alpha \ln(2)}.$$ 
		For example for $\alpha=1$ and $h=256$, the optimal chunk size is equal to \emph{369 bits}.
		Therefore, the order of data that must be uploaded to the blockchain for the disputation phase is reduced from $O(N)$ to $O(\log(N))$.
		
\subsubsection{Computation Load of the Smart Contract}		
Moreover, the computation load of the smart contract in the disputation phase is also reduced to $O(\log(N))$. The reason is that the smart contract verifies a Merkle proof, in Step~\ref{cost1_2} of Algorithm~\ref{algorithm2_2}, with $O(\log(N))$ computation load and computes the $\hash$ of the decryption of an encrypted chunk of the data, in Step~\ref{cost2_2} of Algorithm~\ref{algorithm2_2}, with $O(1)$ computation load.

\subsubsection{Privacy of The Data}		
Similar to the previous algorithm, here also the privacy of the data is perfectly guaranteed. However, in addition to that,  in the disputation phase, at most one chunk of $\overline{D_T}$ is revealed.

\section{Platform Improvement: $O(1)$-Algorithm}\label{optimization3}
\begin{figure}[btp]
			\centering
			\includegraphics[width=90mm,height=10cm,keepaspectratio]{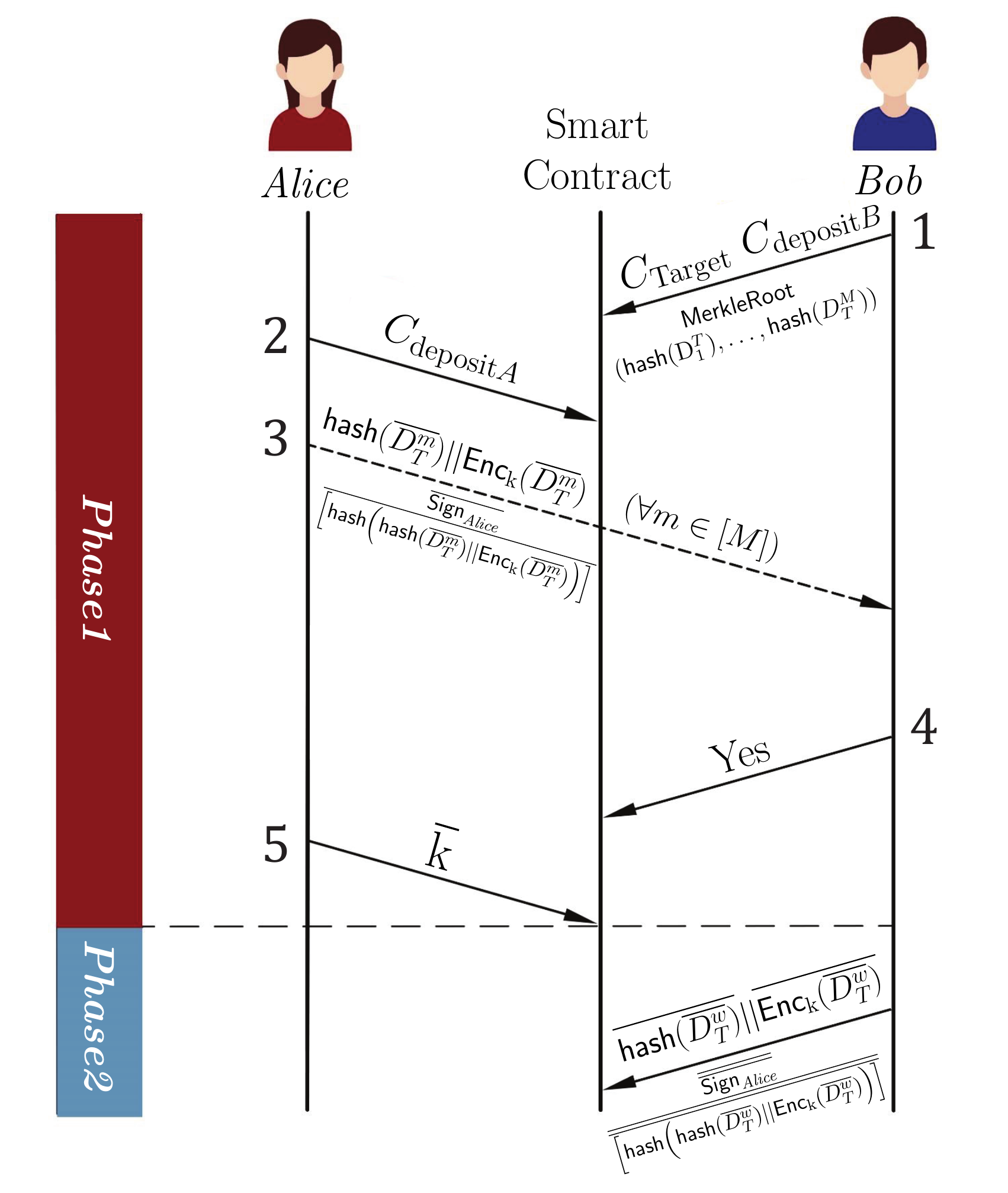}
			\caption{The messages transferred in $O(1)$-Algorithm}
			\label{fig_Optimization3}
			\vskip -0.2in
		\end{figure}

Recall that in the disputation phase of both $O(N)$-Algorithm and $O(\log(N))$-Algorithm, \emph{Bob} needs to prove that the data that he sends to the smart contract, in the disputation phase,  is the same as what he has received from \emph{Alice} through the off-chain channel. This proof requires upload cost of, respectively, $O(N)$ and $O(\log(N))$ in $O(N)$-Algorithm and $O(\log(N))$-Algorithm. In this section, we propose an alternative approach to reduce this cost  $O(1)$ using \emph{Alice}'s signature. In this approach, \emph{Alice} sends to \emph{Bob}, the $\hash$ of each chunk of the data, the encrypted version of each chunk, along with her signature of those contents, through the off-chain channel. Since the public key of \emph{Alice} is already available on the smart contract, in case that \emph{Bob}  wants to dispute the validity of one chunk, he can very easily prove that he received that chunk from \emph{Alice}, as detailed in  Algorithm~\ref{algorithm1_3} (Figure~\ref{fig_Optimization3}). 
		
		\begin{algorithm}[btp]
			\caption{$O(1)$-Algorithm: Trading Phase}
			\label{algorithm1_3}
			\begin{algorithmic}[1]
	\Procedure {Initialization}{}	
			\State \emph{Alice} owns data $D_T$, which is divided into $M \in \mathbb{N}$ chunks, each of size $L \in \mathbb{N}$ bits, as $D_T = (D_T^1,\ldots,D_T^M)$. Data $D_T$ is certified by \emph{Carol}. This means \emph{Carol} has uploaded $\MerkleRoot(\hash(D_T^1),\ldots,\hash(D_T^M))$ and $\Sign_{\emph{Carol}}\Big[\MerkleRoot\big(\hash(D_T^1),\ldots,\hash(D_T^M)\big)\Big]$ to the smart contract (see Figure~\ref{fp11a}).
\State \emph{Alice} and \emph{Bob} agree upon price $C_{\textrm{Target}}$, and also $C_{\textrm{deposit}A}$ and $C_{\textrm{deposit}B}$.
\State \emph{Alice} generates a key, denoted by \Key. 	
\EndProcedure
	\Statex
\Procedure {Trading Phase}{}
				\State{\emph{Bob} sends $\MerkleRoot(\hash(D_T^1),\ldots,\hash(D_T^M))$ to the smart contract, showing his interest in buying $D_T$.}
				\State{\emph{Bob} deposits $C_{\textrm{Target}}$ and $C_{\textrm{deposit}B}$ to the smart contract, nonrefundable for one day.}
				\State{\emph{Alice} deposits $C_{\textrm{deposit}A}$ to the smart contract, nonrefundable for one day.}
				
				\Comment{If Algorithm doesn't get to Step~\ref{sendK_3} until the end of one day, it will go to Step~\ref{refundable_3}.}

				\State{\emph{Alice} generates $\Enc_\Key(\overline{D_T^m}),\forall m \in [M],$ as an encrypted version of $\overline{D_T^m}$, using key $\Key$, where $\overline{D_T^m}$ is the $m^{th}$ chunk of data $\overline{D_T}$. \emph{Alice} sends $\hash(\overline{D_T^m}) || \Enc_\Key(\overline{D_T^m})$, $\forall m \in [M]$, to \emph{Bob} through the off-chain channel. Moreover, she sends the signed of $\hash$ of their concatenation using her private key (the one paired with her public key on the blockchain), $\overline{\Sign_{\emph{Alice}}\Big[\hash\Big(\hash(\overline{D_T^m}) || \Enc_\Key(\overline{D_T^m})\Big)\Big]}$ 
				to \emph{Bob} through the off-chain channel.}\label{sendingencK_3}
				\State{\emph{Bob} checks if:
				\begin{enumerate}[label=\alph*]
					\item the signature of $\emph{Alice}$ in $\overline{\Sign_{\emph{Alice}}\Big[\hash\Big(\hash(\overline{D_T^m}) || \Enc_\Key(\overline{D_T^m})\Big)\Big]}$  is verified, using  $\emph{Alice}$ public key and  $\hash(\overline{D_T^m}) || \Enc_\Key(\overline{D_T^m})$, $\forall m \in [M]$. 
					\item $\MerkleRoot(\hash(\overline{D_T^1}),\ldots,\hash(\overline{D_T^M})) == \MerkleRoot(\hash(D_T^1),\ldots,\hash(D_T^M))$
				\end{enumerate}
				and sends "Yes" to the smart contract, if the above equalities hold. If \emph{Bob} sends "No" to the smart contract or remains silent until the end of one day, the smart contract goes to Step~\ref{refundable_3}}.\label{sendYes_3}
				
				\State{After receiving "Yes" by the smart contract from \emph{Bob}, \emph{Alice} sends $\overline{\Key}$ to the smart contract, claiming it is indeed key $\Key$.}\label{sendK_3}
				\State{\emph{Bob} can check validity of the received data by decrypting each encrypted chunk using key $\overline{\Key}$ (available on the smart contract), then computing the $\hash$ of each decrypted chunk, and then comparing each with the version that \emph{Alice} sent to him through the P2P channel. In other words, he can check the following equality for all chunks:
						$$\hash\Big(\Dec_{\overline{\Key}}\big(\Enc_\Key(\overline{D_T^m})\big)\Big) == \hash\big(\overline{D_T^m}\big), m \in [M]$$
						If the above equality is not valid for at least one chunk, say chunk $w \in [M]$, \emph{Bob} sends $\Enc_\Key(\overline{D_T^w}) || \hash(\overline{D_T^w})$, and $\overline{\Sign_{\emph{Alice}}\Big[\hash\Big(\hash(\overline{D_T^w}) || \Enc_\Key(\overline{D_T^w})\Big)\Big]}$  to the smart contract. Otherwise, he does not send anything to the smart contract.} \label{Bob_hash_Dec_3}
				\If {the smart contract receives no objection from \emph{Bob} in a determined grace period (say 2 days)}
				\State {$C_{\textrm{Target}}$ is refundable to \emph{Alice}, and $C_{\textrm{deposit}A}$ and $C_{\textrm{deposit}B}$ to \emph{Alice} and \emph{Bob} respectively.}
				\State{The algorithm terminates.}
				\Else
				\State Go to Disputation Phase (Algorithm~\ref{algorithm2_3}).
				\EndIf
				\State{Deposits be refundable.
					
					$C_{\textrm{Target}}$ and $C_{\textrm{deposit}B}$ are refundable to \emph{Bob} and $C_{\textrm{deposit}A}$ to \emph{Alice}. Algorithm terminates.}
				\label{refundable_3}
				\EndProcedure
			\end{algorithmic}
		\end{algorithm}

		\begin{algorithm}[H]
			\caption{$O(1)$-Algorithm: Disputation Phase}
			\label{algorithm2_3}
			\begin{algorithmic}[1]
			\Procedure {Disputation Phase}{}
				\State{If there is a chunk $w \in [M]$, such that $\hash\Big(\Dec_{\overline{\Key}}\big(\Enc_\Key(\overline{D_T^w})\big)\Big) \neq \hash\big(\overline{D_T^w}\big)$, \emph{Bob} sends its $\hash$ and encrypted version, i.e., $\overline{\hash(\overline{D_T^w})} || \overline{\Enc_\Key(\overline{D_T^w})}$, along with 
				$\overline{\overline{\Sign_{\emph{Alice}}\Big[\hash\Big(\hash(\overline{D_T^w}) || \Enc_\Key(\overline{D_T^w})\Big)\Big]}}$
 to the smart contract, claiming he received it from \emph{Alice} in Step~\ref{sendingencK_3} of Algorithm~\ref{algorithm1_3}.}
				
				\Comment{Recall that \emph{Bob} received these data from \emph{Alice} in Step~\ref{sendingencK_3} of Algorithm~\ref{algorithm1_3}}.
				
				\State{Using the uploaded data by \emph{Bob}, at first, the smart contract checks signature of data.}
				
				\If{ \emph{Alice}'s signature in $\overline{\overline{\Sign_{\emph{Alice}}\Big[\hash\Big(\hash(\overline{D_T^w}) || \Enc_\Key(\overline{D_T^w})\Big)\Big]}}$ is verified using $\hash\Big(\overline{\hash(\overline{D_T^w})} || \overline{\Enc_\Key(\overline{D_T^w})}\Big)$}			
			
				\State{Using key $\overline{\Key}$, the smart contract calculates $\Dec_{\overline{\Key}}\big(\overline{\Enc_\Key(\overline{D_T^w})}\big)$  as decryption of $\overline{\Enc_\Key(\overline{D_T^w})}$, and then computes $\hash\Big(\Dec_{\overline{\Key}}\big(\overline{\Enc_\Key(\overline{D_T^w})}\big)\Big)$.}
				
				\Comment{ Recall that the smart contract received $\overline{\Key}$ from \emph{Alice} in Step~\ref{sendK_3} of Algorithm~\ref{algorithm1_3}.}
				
				\If {$\hash\Big(\Dec_{\overline{\Key}}\big(\overline{\Enc_\Key(\overline{D_T^w})}\big)\Big) == \overline{\hash(\overline{D_T^w})}$}
				\State{go to Step~\ref{badBob_3}: 
					\emph{Bob} is dishonest.}
				\Else
				\State{ go to Step~\ref{badAlice_3}: 
					\emph{Alice} is dishonest.}
				\EndIf
				\Else
				\State{ go to Step~\ref{badBob_3}: 
					\emph{Bob} is dishonest.}
				\EndIf

				\State{\emph{Alice} is dishonest.
					
					$C_{\textrm{Target}}$, $C_{\textrm{deposit}A}$, and $C_{\textrm{deposit}B}$ are refundable to \emph{Bob}. Algorithm terminates.}

				\label{badAlice_3}
				
				\State{\emph{Bob} is dishonest.
					
					$C_{\textrm{Target}}$, $C_{\textrm{deposit}A}$, and $C_{\textrm{deposit}B}$ are refundable to \emph{Alice}. Algorithm terminates. }				
				\label{badBob_3}
\EndProcedure				
			\end{algorithmic}
		\end{algorithm}
		
		\subsection{Analysis of $O(1)$-Algorithm}
		It is easy to verify that the first to fifth requirements are satisfied in this algorithm. Also, it is easy to see that the computation and storage cost of the trading phase to the smart contract is constant. Here we want to evaluate the computation load for the parties and the computation and storage cost of the disputation phase to the smart contract.
		\subsubsection{Computation Load for the Parties}
		According to Algorithm~\ref{algorithm1_3}, the computation load for \emph{Alice} is computing the encryption of each chunk of the data and the signature of the $\hash$ of $\hash(\overline{D_T^m}) || \Enc_\Key(\overline{D_T^m})$ (Step~\ref{sendingencK_3}), and for \emph{Bob} is computing verification of the signature of \emph{Alice} in the $\hash$ of $\hash(\overline{D_T^m}) || \Enc_\Key(\overline{D_T^m})$, $\MerkleRoot(\hash(\overline{D_T^1}),\ldots,\hash(\overline{D_T^M}))$, and $\hash\Big(\Dec_{\overline{\Key}}\big(\Enc_\Key(\overline{D_T^m})\big)\Big), \forall m \in [M]$ (Steps~\ref{sendYes_3} and \ref{Bob_hash_Dec_3}). Let us assume that for any data $D$, $\frac{\C(\Sign(D))}{|D|}=\beta_4$, for some $\beta_4 > 0$. Similar to the analysis of the previous algorithm, the computation load for the parties is $O(N)$, where $|D_T|=N$.
		
		\subsubsection{Size of the Data Uploaded to the Blockchain}
		
Note that the volume of the disputation data in this algorithm is constant, so by considering 65 bytes length for the signature~\cite{solidity2019introduction}, $h=256$, $\alpha=1$, and $L=256$ the volume of the data needed for the disputation is 1032 bits. Therefore, the order of data that must be uploaded to the blockchain for the disputation is reduced to $O(1)$. 
		
\subsubsection{Computation Load of the Smart Contract}		
Also, the computation load of the smart contract in the disputation phase is also $O(1)$, which is very desirable.

\subsubsection{Privacy of The Data}		
In this approach also, in the disputation phase, only one chunk of $\overline{D_T}$  is revealed. 
	
\section{Conclusion} \label{sec_Conclusion}
	In this paper, by exploiting the advantages of blockchain and smart contracts, we propose \myname \space platform as a decentralized data market that does not require any mutual trust between the trade parties or a trusted third  party as a mediator.  In the proposed platform, the computation and storage load on the smart contract is negligible and constant if the parties behave honestly. In the presence of malicious behavior, the proposed algorithm allows the honest party to prove the malicious behavior of the other party to the smart contract, again with $O(1)$ computation and storage cost to the blockchain.

\bibliography{myref}
\bibliographystyle{ieeetr}

\end{document}